\documentclass[twocolumn,superscriptaddress,showpacs,nofootinbib]{revtex4-1}

\usepackage[colorlinks,pdfstartview=FitH]{hyperref}
\hypersetup{linkcolor=blue,citecolor=blue,filecolor=black,urlcolor=blue}

\usepackage{amsmath,amssymb,bm}
\usepackage{slashed}
\usepackage{graphicx,tikz}
\usetikzlibrary{positioning}
\usepackage{amsfonts}
\usepackage{color}

\usepackage{amsfonts}
\usepackage{bbm}
\usepackage{latexsym}
\allowdisplaybreaks[1]

\begin{document}

\title{Tamed loops: A try for non-renormalizable Einstein gravity in UV-free scheme}

\author{Lian-Bao Jia}  \email{jialb@mail.nankai.edu.cn}
\affiliation{
School of Mathematics and Physics, Southwest University of Science and Technology, Mianyang 621010, China}
\affiliation{Department of Physics, Chongqing Key Laboratory for Strongly Coupled Physics, Chongqing University, Chongqing 401331, China}

\begin{abstract}

How to describe loop corrections of gravitation is a fundamental challenge in the quantization of Einstein gravity. In this paper, we give it a try in UV-free scheme, including one-loop propagator and two-loop vertex, and the results are effective for graviton loops. This indicates that both loops of the renormalizable Standard Model and the non-renormalizable Einstein gravity can be described in a unified way.

\end{abstract}


\maketitle

\section{Introduction}

There are four fundamental interactions (the electromagnetic interaction, the weak and strong interactions, and the gravitational interaction) currently known to exist in nature, and three of them (excluding gravity) are described within the framework of quantum field theory, i.e., the Standard Model (SM) of particle physics. Today, the widely-accepted and well-tested theory of gravity is Einstein's general theory of relativity (GR) \cite{Einstein:1915ca}, which is considered as an effective gravitational theory below the Planck scale. When one tries to quantize the classical field of Einstein gravity, an insurmountable obstacle appears --- the non-renormalizability of gravity (due to the negative mass dimension of the coupling coefficient).

Let's give some brief explications about the renormalizability of a theory. For loop corrections in quantum field theory, the results are often ultraviolet (UV) divergences when one evaluates the integration over free momenta in loops. To make sense of UV divergences and extract finite results from infinities, a paradigm approach is regularization (such as Pauli-Villars regularization \cite{Pauli:1949zm}, dimensional regularization \cite{tHooft:1972tcz}) with divergences mathematically expressed and renormalization with divergences removed by counterterms, i.e. divergences mathematically removed by $\infty - \infty$. This paradigm depends on the Bogoliubov-Parasiuk-Hepp-Zimmermann (BPHZ) renormalization scheme \cite{Bogoliubov:1957gp}, and there are a finite number of counterterms needed during renormalization in a renormalizable theory. For a non-renormalizable theory with negative-mass-dimension coupling coefficients, such as GR, it requires an infinite number of counterterms to cure all UV divergences of loops. The theory of Einstein gravity is non-renormalizable \cite{DeWitt:1967yk,tHooft:1974toh,Goroff:1985th}.

How to describe loop corrections of gravitation? This is an open fundamental challenge in modern physics.  A variety of approaches are explored to describe possible quantum behavior of the gravitational field, and two popular approaches are string theory/M-theory and loop quantum gravity. In this paper, we focus on the quantization of Einstein gravity, and other types of gravity are beyond the scope of this paper. Now, the question is more specific: Is there an effective method to describe the loops of non-renormalizable GR? In above discussions, proper regulators with BPHZ scheme can be adopted to the quantization of three fundamental interactions in SM but with gravity excluded.

\begin{figure}[htbp!]
\includegraphics[width=0.45\textwidth]{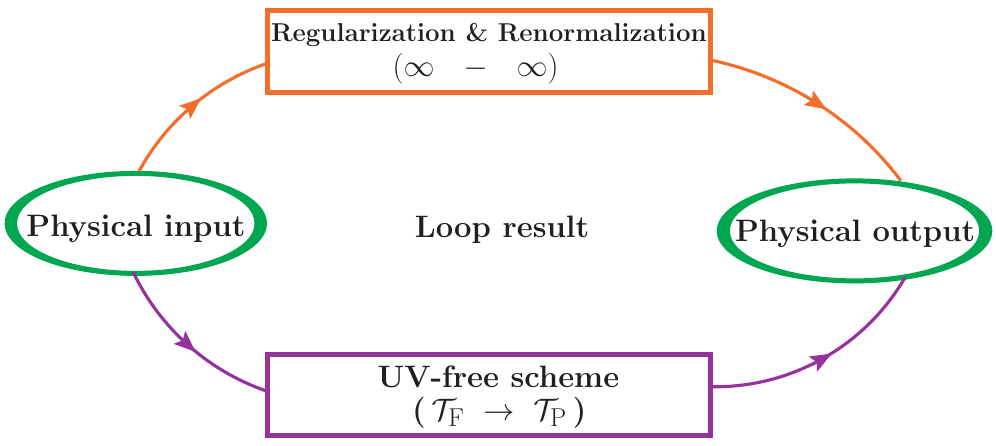}
\caption{A schematic diagram of the loop result derived by two routes of concern.}
\label{loop-re}
\end{figure}

Besides renormalization for logarithmic divergences, further changes to the UV divergence description of loops are required before a possible general approach, as pointed out by Dirac \cite{DiracTheEO}. If we go forward in the direction pointed out by Dirac, i.e., adopt a new proper method to deal with UV divergences of loops (both logarithmic and power-law divergences), then there may be a way out of the dilemma in the quantization of gravity. Here we pay attention to an effective method of the UV-free scheme \cite{Jia:2023dub}, which is one of the alternative routes between the physical input and output, as depicted in Fig. \ref{loop-re}:\\
(a) \textit{Equivalent transformation} of the loop integral, regularization \& renormalization ($\infty - \infty$). \\
(b) \textit{Analytic continuation} of the transition amplitude, the UV-free scheme ($\mathcal{T}_\mathrm{F} \to  \mathcal{T}_\mathrm{P}$). \\
In UV-free scheme, the loop contribution is assumed to a local contribution with contributions from UV regions being insignificant, and the finite loop result can be obtained without UV divergences, i.e. our lack of high-energy behaviors doesn't seem to prevent us from effectively extracting local corrections at low-energy scale. Since there is no UV divergence in calculations, besides its applications to the renormalizable fields in SM \cite{Jia:2023dub}, it may also be capable to describe loop contributions of the non-renormalizable Einstein gravity below the Planck scale. We will try it in this paper.

\section{Action}

The Einstein-Hilbert action is
\begin{eqnarray}
\mathcal{S}_{\mathrm{EH}} \!=\!  \int d^4 X \sqrt{-g}\frac{2}{\kappa^2} (-R-2\Lambda)  \, ,
\end{eqnarray}
with $\kappa = \sqrt{32 \pi G}$ adopted (here the metric of the flat Minkowski spacetime is $\eta_{\mu \nu}$ = diag($1,-1,-1,-1$)). The cosmological constant $\Lambda$ is negligible at ordinary scales. With a matter field term $\mathcal{L}_\mathrm{M}$ added, the action is
\begin{eqnarray}
\mathcal{S} \!=\!  \int d^4 X \sqrt{-g}\big[-\frac{2}{\kappa^2} R + \mathcal{L}_\mathrm{M} \big]\, ,
\end{eqnarray}
which yields the Einstein field equations when one takes the variation $\delta g^{\mu\nu}$ of this action. In a weak field expansion with a small fluctuation of the metric $g_{\mu \nu}$ around a flat background of Minkowski spacetime $\eta_{\mu \nu}$, the metric field can be written as
\begin{eqnarray}
g_{\mu \nu} =\eta_{\mu \nu} + \kappa h_{\mu \nu} \, ,
\end{eqnarray}
with $h_{\mu \nu}$ the quantum fluctuations. The perturbation of $h_{\mu \nu}$ field can describe a gauge theory of massless spin-2 graviton, which will reduce to Einstein gravity at large distances \cite{Weinberg:1965rz}. Adding a gauge fixing term $\mathcal{L}_{gf}^0$ and the ghost field term $\mathcal{L}_{ghost}^0$ to the gravitational Lagrangian $\mathcal{L}^0$, the result is \cite{Kalmykov:1998cv,Prinz:2020nru,Jakobsen:2020ksu,SevillanoMunoz:2022tfb,Latosh:2022ydd,Latosh:2023zsi}
\begin{eqnarray}
\mathcal{L}^0= -\frac{2}{\kappa^2}\sqrt{-g}R + \mathcal{L}_{gf}^0 + \mathcal{L}_{ghost}^0 + \sqrt{-g} \mathcal{L}_\mathrm{M} \, ,
\end{eqnarray}
and the above Lagrangian is a general form. In the harmonic gauge, the condition is
\begin{eqnarray}
G^\mu=g^{\alpha\beta}\Gamma_{\alpha\beta}^\mu=0 ,
\end{eqnarray}
and a gauge fixing term is
\begin{eqnarray}
\mathcal{L}_{gf}^0= \sqrt{-g} \frac{\zeta}{2\kappa^2} g_{\mu\nu} G^\mu G^\nu\, ,
\end{eqnarray}
where $\zeta$ is a gauge fixing parameter. The Lagrangian of the ghost fields $c$ and $\bar{c}$ is
\begin{eqnarray}
\mathcal{L}_{ghost}^0&=& \sqrt{-g} (- g^{\mu\nu} g^{\alpha\beta} \nabla_\alpha \bar{c}_\mu \nabla_\beta c_\nu  \\ \nonumber
&& - 2 \Gamma_{\alpha\beta}^\mu \bar{c}_\mu \nabla^\alpha c^\beta  + R_{\mu\nu} \bar{c}^\mu c^\nu)       \, .
\end{eqnarray}
Now, the total action is
\begin{eqnarray}
\mathcal{S}^0 \!=\!  \int d^4 X \, \mathcal{L}^0\, .   \label{grav-act-g}
\end{eqnarray}
The Feynman rules for gravitation (see Refs. \cite{Prinz:2020nru,Jakobsen:2020ksu,SevillanoMunoz:2022tfb,Latosh:2022ydd,Latosh:2023zsi}) are commonly obtained in the background field expansion of the Lagrangian $\mathcal{L}^0$ (the extraction of $h_{\mu \nu}$ terms). Taking the parameter $\zeta = 2$, the propagator of graviton is in a simple form,
\begin{eqnarray}
\frac{i\Pi_{\mu\nu\alpha\beta}/2}{p^2 +i\epsilon} \, ,
\end{eqnarray}
with
\begin{eqnarray}
\Pi_{\mu\nu\alpha\beta}=\eta_{\mu\alpha}\eta_{\nu\beta}+\eta_{\mu\beta}\eta_{\nu\alpha}-\eta_{\mu\nu}\eta_{\alpha\beta} \, .
\end{eqnarray}

Let us take a look at the action Eq. (\ref{grav-act-g}) in the view of a general coordinate transformation. In a weak field expansion of the Lagrangian $\mathcal{L}^0$, the coordinate volume element $d^4 X$ is not invariant under a general coordinate transformation. If the gravitational Lagrangian $\mathcal{L}^0$ is written as
\begin{eqnarray}
\mathcal{L}^0 = \sqrt{-g} \mathcal{L} \, ,
\end{eqnarray}
with $\mathcal{L}$ a reduced gravitational Lagrangian, and hence the action Eq. (\ref{grav-act-g}) becomes
\begin{eqnarray}
\mathcal{S}^0 \!=\!  \int d^4 X  \sqrt{-g} \mathcal{L}\, .
\end{eqnarray}
The coordinate invariant volume element $d^4 X \sqrt{-g}$ is restored. Here the reduced gravitational Lagrangian is
\begin{eqnarray} \label{grav-l}
\mathcal{L}= -\frac{2}{\kappa^2} R + \mathcal{L}_{gf} + \mathcal{L}_{ghost} + \mathcal{L}_\mathrm{M} \, ,
\end{eqnarray}
with
\begin{eqnarray}
\mathcal{L}_{gf}= \frac{\mathcal{L}_{gf}^0}{\sqrt{-g}} = \frac{\zeta}{2\kappa^2} g_{\mu\nu} G^\mu G^\nu\, ,
\end{eqnarray}
\begin{eqnarray}
\mathcal{L}_{ghost}= \frac{\mathcal{L}_{ghost}^0}{\sqrt{-g}} &=&  - g^{\mu\nu} g^{\alpha\beta} \nabla_\alpha \bar{c}_\mu \nabla_\beta c_\nu  \\ \nonumber
&& - 2 \Gamma_{\alpha\beta}^\mu \bar{c}_\mu \nabla^\alpha c^\beta  + R_{\mu\nu} \bar{c}^\mu c^\nu       \, .
\end{eqnarray}
In a weak field expansion, a metric $\bar{g}_{\mu \nu}$ has a small fluctuation of $\kappa h_{\mu \nu}$ and is shifted to a metric $g_{\mu \nu}$, with
\begin{eqnarray}
g_{\mu \nu} = \bar{g}_{\mu \nu} + \kappa h_{\mu \nu} \, ,
\end{eqnarray}
and an action with a metric fluctuation is
\begin{eqnarray} \label{grav-relation}
\mathcal{S}_h = \int d^4 \bar{X}  \sqrt{-\bar{g}} \, \mathcal{L}(g_{\mu \nu})  \, ,
\end{eqnarray}
where the coordinate invariant volume element is transformed into that of the background metric $\bar{g}_{\mu \nu}$, i.e. the $d^4 \bar{X} \sqrt{-\bar{g}}$, and $\mathcal{L}(g_{\mu \nu})$ is the form of Lagrangian Eq. (\ref{grav-l}) with $g_{\mu \nu} = \bar{g}_{\mu \nu} + \kappa h_{\mu \nu}$ adopted in the expansion around $\bar{g}_{\mu \nu}$ (indices are raised and lowered by the background metric $\bar{g}_{\mu \nu}$). Supposing that graviton field $h_{\mu \nu}$ can be described by the action with metric fluctuation in a weak field expansion, it means that we can just take the expansion of the reduced gravitational Lagrangian $\mathcal{L}$ in forms of the quantum field $h_{\mu \nu}$ around the metric $\bar{g}_{\mu \nu}$, with the coordinate invariant volume transformation $d^4 X  \sqrt{-g}$ $\to d^4 \bar{X}  \sqrt{-\bar{g}}$.

In the following of this paper, the graviton field described by Eq. (\ref{grav-relation}) is adopted in quantizing Einstein gravity, i.e. gravitation is considered as the fluctuation of the background metric $\bar{g}_{\mu \nu}$. In the case that the spacetime is locally flat, the Minkowski spacetime can be taken as the background spacetime. More specifically, for a weak field expansion around Minkowski spacetime with $\bar{g}_{\mu \nu} = \eta_{\mu \nu}$, one has $g_{\mu \nu} =\eta_{\mu \nu} + \kappa h_{\mu \nu}$, and the action of Eq. (\ref{grav-relation}) becomes
\begin{eqnarray} \label{grav-M}
\mathcal{S}_h = \int d^4 x  \, \mathcal{L}(g_{\mu \nu})    \, ,
\end{eqnarray}
where $d^4 x$ is the volume element in Minkowski spacetime. The expansion of the reduced gravitational Lagrangian $\mathcal{L}$ in the graviton field $h_{\mu \nu}$ can be adopted to describe the corresponding quantum gravity. If possible metric fluctuation is negligible, the action Eq. (\ref{grav-M}) will regress to the familiar form in quantum field theory in Minkowski spacetime with this smooth transition.

\section{Loop correction}

As mentioned in the Introduction, the insurmountable obstacle in the quantum field description of Einstein gravity is the non-renormalizability of gravity in paradigmatic loop calculations. Here we focus on pure gravity in a weak field expansion around Minkowski spacetime described by the action of Eq. (\ref{grav-M}). It is a common case for graviton loops with power-law divergences (e.g. quartic and quadratic divergences), and here we try to describe graviton loops in UV-free scheme. The Feynman rules of multi-graviton are listed in the Appendix A. In the following, we will first evaluate the one-loop propagator of graviton as an application, then turn to the case of $n-$loop graviton corrections with overlapping divergences.

\subsection{One-loop propagator}

\begin{figure}[htbp!]
\includegraphics[width=0.45\textwidth]{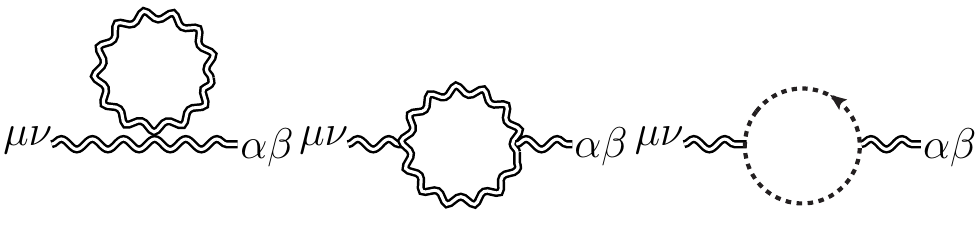} \vspace*{-1ex}
\caption{The one-loop diagrams of graviton propagator.}
\label{one-loop}
\end{figure}

Let's first pay attention to the one-loop propagator of graviton, with processes shown in Fig. \ref{one-loop}. To make the presentation appear complete, the UV-free scheme is briefly listed below. In this scheme, physical contributions of loops are assumed to be local corrections related to external momenta and particle masses with UV regions being insignificant, and the physical transition amplitude $\mathcal{T}_\mathrm{P}$ of loops can be described by an equation \cite{Jia:2023dub}
\begin{eqnarray}
\!\!\!\!\! \mathcal{T}_\mathrm{P} \! =\! \bigg[\! \int \! d\xi_1 \cdots d\xi_i \frac{\partial\mathcal{T}_\mathrm{F}(\xi_1, \cdots, \xi_i)}{\partial\xi_1 \cdots \partial\xi_i}   \!\bigg]_{\!\{\xi_1, \cdots, \xi_i\}\rightarrow0} \!+\! C  ,
\end{eqnarray}
or an equivalent form
\begin{eqnarray}
\!\!\!\!\! \mathcal{T}_\mathrm{P} \! =\! \bigg[\! \int \! (d\xi)^n  \frac{\partial^n\mathcal{T}_\mathrm{F}(\xi)}{\partial\xi^n}   \!\bigg]_{\xi \to 0} \!+\! C  ,
\end{eqnarray}
where the Feynman-like amplitude $\mathcal{T}_\mathrm{F}(\xi_1, \cdots, \xi_i)$ is written by Feynman rules, with parameters $\xi_1, \cdots, \xi_i$ added into denominators of propagators, and $\int \! (d\xi)^n$ means $n$-times antiderivative with respect to $\xi$. For loops with UV divergence inputs, the evaluation first is the loop momentum and then is the $\xi$ parameter (they are noncommutative for loops with UV divergence inputs). The primary antiderivative (expressions of the $\Big[ \cdots \Big]$) is the core in describing the physical transition amplitude, and a rule ($\xi-$dependent choice) for the primary antiderivative is introduced in the Appendix B. For a UV divergent loop with the divergence to a power of 2$n'$, if a UV-convergence transformation $ \bigg[\! \int \! (d\xi)  \frac{\partial\mathcal{T}_\mathrm{F}(\xi)}{\partial\xi} \!\bigg]_{\xi \to 0}$ is denoted as $\mathcal{T}_\mathrm{F}^{[1]} (\xi)$, we have a concise form $\mathcal{T}_\mathrm{P} = \mathcal{T}_\mathrm{F}^{[n]} (\xi) +C$ with $n \geq n'+1$, i.e. $\mathcal{T}_\mathrm{P}$ has UV transformation invariance and is the analytic continuation $\mathcal{T}_\mathrm{F} \! \to \! \mathcal{T}_\mathrm{P}$ in physics.

The Feynman-like transition amplitude $\mathcal{T}_\mathrm{F}^{a} (\xi_1, \xi_2, \xi_3)$ in the first diagram of Fig. \ref{one-loop} can be written as
\begin{eqnarray}
\mathcal{T}_\mathrm{F}^{a} (\xi_1, \!\xi_2,\! \xi_3) \!&=&\!\frac{2 i \kappa^2}{2} \!\! \mathrm{\int} \!\! \frac{d^4k}{(2\pi)^4} \frac{i\Pi_{\mu_3\nu_3\mu_4\nu_4}/2}{k^2 \!+\!\xi_1\!+\! \xi_2 \!+\!\xi_3 } \\
&& \times \Big(V^{\mu_3\nu_3\mu_4\nu_4|\lambda_1\mu\nu \lambda_2\alpha\beta} (p)_{\lambda_1} (-p)_{\lambda_2}   \nonumber \\
&&+ V^{\alpha\beta\mu_4\nu_4|\lambda_1\mu\nu \lambda_3\mu_3\nu_3} (p)_{\lambda_1} (-k)_{\lambda_3} \nonumber\\
&&+V^{\alpha\beta\mu_3\nu_3|\lambda_1\mu\nu \lambda_4\mu_4\nu_4} (p)_{\lambda_1} (k)_{\lambda_4} \nonumber\\
&&+V^{\mu\nu \mu_4\nu_4|\lambda_2\alpha\beta \lambda_3\mu_3\nu_3} (-p)_{\lambda_2} (-k)_{\lambda_3} \nonumber \\
&&+V^{\mu\nu\mu_3\nu_3|\lambda_2\alpha\beta \lambda_4\mu_4\nu_4} (-p)_{\lambda_2} (k)_{\lambda_4} \nonumber \\
&&+V^{\mu\nu\alpha\beta|\lambda_3\mu_3\nu_3 \lambda_4\mu_4\nu_4} (-k)_{\lambda_3} (k)_{\lambda_4}\Big) \nonumber \, .
\end{eqnarray}
The physical transition amplitude $\mathcal{T}_\mathrm{P}^{a}$ is
\begin{eqnarray}
\mathcal{T}_\mathrm{P}^{a}\! & = & \! \Big [ \mathrm{\int} d\xi_1 d\xi_2 d\xi_3 \frac{\partial\mathcal{T}_\mathrm{F}^{a} (\xi_1, \!\xi_2,\! \xi_3)}{\partial\xi_1 \partial\xi_2 \partial\xi_3}  \Big]_{\{\xi_1, \xi_2, \xi_3\} \to 0} \! + \! C_a^{\mu\nu\alpha\beta}   \nonumber    \\
&=& \! \Big [ \frac{2 i \kappa^2}{2}  \! \! \mathrm{\int} \! d\xi_1 d\xi_2 d\xi_3  \! \mathrm{\int} \! \frac{d^4k}{(2\pi)^4} \frac{(-3!) i\Pi_{\mu_3\nu_3\mu_4\nu_4}/2}{(k^2 \!+\!\xi_1\!+\! \xi_2 \!+\!\xi_3 )^4}   \\
&& \times \Big(V^{\mu_3\nu_3\mu_4\nu_4|\lambda_1\mu\nu \lambda_2\alpha\beta} p_{\lambda_1}^{} (-p)_{\lambda_2}^{}    \nonumber \\
&&+V^{\mu\nu\alpha\beta|\lambda_3\mu_3\nu_3 \lambda_4\mu_4\nu_4} (-k)_{\lambda_3}^{} k_{\lambda_4}^{} \! \Big)  \Big]_{\{\xi_1, \xi_2, \xi_3\} \to 0} \!\! + \! C_a^{\mu\nu\alpha\beta} \nonumber \, .
\end{eqnarray}
It is UV-converged when evaluating the loop momentum $k$. After integral, the result is
\begin{eqnarray}
\mathcal{T}_\mathrm{P}^{a} \!
&=&  \! \Big [ i \kappa^2   \frac{i\Pi_{\mu_3\nu_3\mu_4\nu_4}}{2} \frac{i}{16\pi^2} \Big( V^{\mu_3\nu_3\mu_4\nu_4|\lambda_1\mu\nu \lambda_2\alpha\beta} p_{\lambda_1}^{} p_{\lambda_2}^{} \nonumber \\
&& \times  (\xi_1 \!-\! \xi_1 \log \xi_1) \!+\! \frac{V^{\mu\nu\alpha\beta|\lambda_3\mu_3\nu_3 \lambda_4\mu_4\nu_4} \eta_{\lambda_3 \lambda_4}^{}}{4}  \\
&& \times (\xi_1^2 \!\log \xi_1\!-\!\frac{3}{2}\xi_1^2) \! \Big)\!\Big]_{\!\xi_1 \to 0}\!+\! C_a^{\mu\nu\alpha\beta} \, . \nonumber
\end{eqnarray}
In the limit $\xi_1 \to 0$, the primary antiderivative is zero (it is an accurate result for massless particles), and $C_a^{\mu\nu\alpha\beta}=0$ is adopted for the case with the primary antiderivative being zero. The physical transition amplitude is $\mathcal{T}_\mathrm{P}^{a} =$ 0.

The physical transition amplitude $\mathcal{T}_\mathrm{P}^{b}$ in the second diagram of Fig. \ref{one-loop} is
\begin{eqnarray}
\mathcal{T}_\mathrm{P}^{b} & = & \! \Big [ \mathrm{\int} d\xi_1 d\xi_2 d\xi_3 \frac{\partial\mathcal{T}_\mathrm{F}^{b} (\xi_1, \!\xi_2,\! \xi_3)}{\partial\xi_1 \partial\xi_2 \partial\xi_3}  \Big]_{\{\xi_1, \xi_2, \xi_3\} \to 0} \!\! + \! C_b^{\mu\nu\alpha\beta}      \\
&=& \! \Big [  \frac{(2 i \kappa)^2}{2}  \! \! \mathrm{\int} \! d\xi_1 d\xi_2 d\xi_3  \! \mathrm{\int} \! \frac{d^4k}{(2\pi)^4} \frac{(-3!) i\Pi_{\mu_2\nu_2\alpha_3\beta_3}/2}{(k^2 \!+\!\xi_1\!+\! \xi_2 \!+\!\xi_3 \!+\!i\epsilon)^4}   \nonumber \\
&& \times \frac{i\Pi_{\alpha_2\beta_2\mu_3\nu_3}/2}{(k-p)^2+i\epsilon} \Big(V^{\mu_3\nu_3|\lambda_1\mu\nu \lambda_2\mu_2\nu_2} p_{\lambda_1}^{} (-k)_{\lambda_2}^{}   \nonumber  \\
&&+V^{\mu_2\nu_2|\lambda_1\mu\nu \lambda_3\mu_3\nu_3} p_{\lambda_1}^{} (k\!-\!p)_{\lambda_3}^{} \!+\!V^{\mu\nu|\lambda_2\mu_2\nu_2 \lambda_3\mu_3\nu_3}   \nonumber \\
&&\times (-k)_{\lambda_2}^{} (k\!-\!p)_{\lambda_3}^{} \! \Big) \Big(\! V^{\alpha_3\beta_3|\rho_1\alpha\beta \rho_2\alpha_2\beta_2} (-p)_{\rho_1}^{} (p\!-\!k)_{\rho_2}^{}   \nonumber  \\
&&+V^{\alpha_2\beta_2|\rho_1\alpha\beta \rho_3\alpha_3\beta_3} (-p)_{\rho_1}^{} k_{\rho_3}^{} +V^{\alpha\beta|\rho_2\alpha_2\beta_2 \rho_3\alpha_3\beta_3}   \nonumber \\
&&\times (p\!-\!k)_{\rho_2}^{} k_{\rho_3}^{} \Big) \Big]_{\{\xi_1, \xi_2, \xi_3\} \to 0} \! + \! C_b^{\mu\nu\alpha\beta} \nonumber \, .
\end{eqnarray}
After integral, one has
\begin{eqnarray}
\mathcal{T}_\mathrm{P}^{b} \! &=& \!  \frac{(2 i \kappa)^2}{2} \frac{i}{16\pi^2} \!\mathrm{\int}^1_{\!\!\! 0}  \!d x  (-3!) \Pi_{\mu_2\nu_2\alpha_3\beta_3} \Pi_{\alpha_2\beta_2\mu_3\nu_3}     \\
&& \times \frac{1}{4!}\Big\{ \Big( W^{33}  p_{\lambda_1}^{} p_{\rho_1}^{} x(1\!-\!x)(\frac{p^2}{2} \eta_{\lambda_2 \rho_2}^{} \!+\!p_{\lambda_2}^{} p_{\rho_2}^{}) \nonumber    \\
&& +  W^{32} p_{\lambda_1}^{} p_{\rho_1}^{}(1\!-\!x)(\frac{-x p^2}{2}\eta_{\lambda_2 \rho_3}^{} \!+\!(1-x)p_{\lambda_2}^{} p_{\rho_3}^{}) \nonumber   \\
&& +  W^{31}  p_{\lambda_1}^{}x(1\!-\!x)(-(1\!-\!x)p_{\lambda_2}^{} p_{\rho_2}^{} p_{\rho_3}^{} \nonumber\\
&&+ \frac{p^2}{2} [x\eta_{\lambda_2 \rho_3}^{} p_{\rho_2}^{}-(1-x)(\eta_{\lambda_2 \rho_2}^{} p_{\rho_3}^{}+ \eta_{\rho_2\rho_3}^{} p_{\lambda_2}^{})]) \nonumber   \\
&&+ W^{23} p_{\lambda_1}^{} p_{\rho_1}^{} x(x p_{\lambda_3}^{}  p_{\rho_2}^{} - \frac{(1\!-\!x)p^2}{2}\eta_{\lambda_3 \rho_2}^{}) \nonumber   \\
&&+ W^{22} p_{\lambda_1}^{} p_{\rho_1}^{} x(1-x)( p_{\lambda_3}^{}  p_{\rho_3}^{} + \frac{p^2}{2}\eta_{\lambda_3 \rho_3}^{}) \nonumber   \\
&&+ W^{21} p_{\lambda_1}^{} x(1\!-\!x)(-x p_{\lambda_3}^{} p_{\rho_2}^{} p_{\rho_3}^{} \nonumber\\
&&+ \frac{ p^2}{2} [\eta_{\lambda_3 \rho_2}^{} p_{\rho_3}^{}\!-\!x(\eta_{\lambda_3 \rho_2}^{} p_{\rho_3}^{}\!+\! \eta_{\rho_2\rho_3}^{} p_{\lambda_3}^{}\!+\!\eta_{\rho_3\lambda_3}^{} p_{\rho_2}^{})])  \nonumber   \\
&&+ W^{13} x(1-x) p_{\rho_1}^{} (-x p_{\lambda_2}^{} p_{\lambda_3}^{} p_{\rho_2}^{} \nonumber \\
&& + \frac{p^2}{2} [(1-x)\eta_{\lambda_3 \rho_2}^{} p_{\lambda_2}^{} -x(\eta_{\lambda_2 \rho_2}^{} p_{\lambda_3}^{}+ \eta_{\lambda_2 \lambda_3}^{} p_{\rho_2}^{})]) \nonumber   \\
&&+W^{12} x(1-x) p_{\rho_1}^{} (-(1-x)p_{\lambda_2}^{} p_{\lambda_3}^{} p_{\rho_3}^{} \nonumber \\
&&+ \frac{p^2}{2} [x\eta_{\lambda_2 \rho_3}^{} p_{\lambda_3}^{} \!-\!(1\!-\!x)(\eta_{\lambda_3 \rho_3}^{} p_{\lambda_2}^{}\!+\! \eta_{\lambda_2 \lambda_3}^{} p_{\rho_3}^{})]) \nonumber   \\
&&+W^{11}x(1-x) [x(1-x)p_{\lambda_2}^{} p_{\lambda_3}^{} p_{\rho_2}^{} p_{\rho_3}^{}  \nonumber \\
&& +\frac{p^4}{8} x(1\!-\!x) (\eta_{\lambda_3 \rho_2}^{} \eta_{\lambda_2 \rho_3}^{}\!\!+\!\eta_{\lambda_2 \rho_2}^{} \eta_{\lambda_3 \rho_3}^{}\!\!+\!\eta_{\lambda_2 \lambda_3}^{} \eta_{\rho_2 \rho_3}^{})  \nonumber \\
&& +\frac{p^2}{2} (x(1\!-\!x)(\eta_{\rho_2 \rho_3}^{} p_{\lambda_2}^{}p_{\lambda_3}^{}\!\!+\!\eta_{\lambda_2 \rho_2}^{} p_{\lambda_3}^{} p_{\rho_3}^{} \!\! +\!\eta_{\lambda_3 \rho_3}^{} p_{\!\lambda_2}^{} p_{\!\rho_2}^{} \nonumber \\
&&+\eta_{\lambda_2 \lambda_3}^{} p_{\!\rho_2}^{} p_{\!\rho_3}^{}) \!-\!x^2\eta_{\lambda_2 \rho_3}^{} p_{\!\lambda_3}^{} p_{\!\rho_2}^{} \!\!-\!(1\!-\!x)^2 \eta_{\lambda_3 \rho_2}^{} p_{\!\lambda_2}^{} p_{\!\rho_3}^{}) ] \!\Big) \nonumber \\
&&\times \log\!\frac{1}{-p^2 x(1\!-\!x)} \Big\} + C_b^{\mu\nu\alpha\beta} \nonumber \, ,
\end{eqnarray}
with the following notes for simplicity,
\begin{eqnarray}
W^{33} = V^{\mu_3\nu_3|\lambda_1\mu\nu \lambda_2\mu_2\nu_2} V^{\alpha_3\beta_3|\rho_1\alpha\beta \rho_2\alpha_2\beta_2} \, , \nonumber
\end{eqnarray}
\begin{eqnarray}
W^{32} = V^{\mu_3\nu_3|\lambda_1\mu\nu \lambda_2\mu_2\nu_2} V^{\alpha_2\beta_2|\rho_1\alpha\beta \rho_3\alpha_3\beta_3} \, , \nonumber
\end{eqnarray}
\begin{eqnarray}
W^{31} = V^{\mu_3\nu_3|\lambda_1\mu\nu \lambda_2\mu_2\nu_2} V^{\alpha\beta|\rho_2\alpha_2\beta_2 \rho_3\alpha_3\beta_3} \, , \nonumber
\end{eqnarray}
\begin{eqnarray}
W^{23} = V^{\mu_2\nu_2|\lambda_1\mu\nu \lambda_3\mu_3\nu_3}  V^{\alpha_3\beta_3|\rho_1\alpha\beta \rho_2\alpha_2\beta_2} \, , \nonumber
\end{eqnarray}
\begin{eqnarray}
W^{22} = V^{\mu_2\nu_2|\lambda_1\mu\nu \lambda_3\mu_3\nu_3}  V^{\alpha_2\beta_2|\rho_1\alpha\beta \rho_3\alpha_3\beta_3} \, , \nonumber
\end{eqnarray}
\begin{eqnarray}
W^{21} = V^{\mu_2\nu_2|\lambda_1\mu\nu \lambda_3\mu_3\nu_3}  V^{\alpha\beta|\rho_2\alpha_2\beta_2 \rho_3\alpha_3\beta_3}  \, , \nonumber
\end{eqnarray}
\begin{eqnarray}
W^{13} = V^{\mu\nu|\lambda_2\mu_2\nu_2 \lambda_3\mu_3\nu_3}  V^{\alpha_3\beta_3|\rho_1\alpha\beta \rho_2\alpha_2\beta_2}  \, , \nonumber
\end{eqnarray}
\begin{eqnarray}
W^{12} = V^{\mu\nu|\lambda_2\mu_2\nu_2 \lambda_3\mu_3\nu_3}  V^{\alpha_2\beta_2|\rho_1\alpha\beta \rho_3\alpha_3\beta_3}  \, , \nonumber
\end{eqnarray}
\begin{eqnarray}
W^{11} = V^{\mu\nu|\lambda_2\mu_2\nu_2 \lambda_3\mu_3\nu_3}  V^{\alpha\beta|\rho_2\alpha_2\beta_2 \rho_3\alpha_3\beta_3}  \, . \nonumber
\end{eqnarray}
After contraction, the result is
\begin{eqnarray}
\mathcal{T}_\mathrm{P}^{b} \! &=& \!  \frac{(2 i \kappa)^2}{2} \frac{i}{16\pi^2} \!\mathrm{\int}^1_{\!\!\! 0}  \!d x  (-\frac{1}{4}) \Big\{\! \frac{1}{16}[40 x^2 (1\!-\!x)^2 p^{\mu} p^{\nu} p^{\alpha} p^{\beta}   \\
&& + 2 p^2 ((1\!-\!2x)^2 (15x^2 \!-\!15x \!-\!2) (p^{\mu} p^{\nu} \eta^{\alpha \beta} \!+\!p^{\alpha} p^{\beta} \eta^{\mu \nu})  \nonumber  \\
&& + (10x^4 \!-\!20x^3 \!+\!17x^2 \!-\!7x \!+\!2)(p^{\nu} p^{\beta} \eta^{\mu \alpha} \!+\!p^{\mu} p^{\beta} \eta^{\nu \alpha}   \nonumber  \\
&& +p^{\nu} p^{\alpha} \eta^{\mu \beta} \!+\!p^{\mu} p^{\alpha} \eta^{\nu \beta}) ) \!+\!p^4((115x^4 \!-\!230x^3 \!+\!103x^2  \nonumber \\
&& +12x +1)\eta^{\mu \nu} \eta^{\alpha \beta} \!+\!(85x^4\!-170x^3\!+139x^2\!-54x+3)  \nonumber \\
&& \times (\eta^{\mu \alpha} \eta^{\nu \beta} \!+\eta^{\mu \beta} \eta^{\nu \alpha}))]  \log\!\frac{1}{-p^2 x(1\!-\!x)}  \! \Big\} \!+\! C_b^{\mu\nu\alpha\beta} \nonumber \, .
\end{eqnarray}

The physical transition amplitude $\mathcal{T}_\mathrm{P}^{c}$ in the third diagram of Fig. \ref{one-loop} is
\begin{eqnarray}
\mathcal{T}_\mathrm{P}^{c}&=& \! \Big [ \mathrm{\int} d\xi_1 d\xi_2 d\xi_3 \frac{\partial\mathcal{T}_\mathrm{F}^{c} (\xi_1, \!\xi_2,\! \xi_3)}{\partial\xi_1 \partial\xi_2 \partial\xi_3}  \Big]_{\{\xi_1, \xi_2, \xi_3\} \to 0} \! + \! C_c^{\mu\nu\alpha\beta}   \nonumber   \\
&=& \!\Big [ (-1)(i \kappa)^{2} \!\! \mathrm{\int} \! d\xi_1 d\xi_2 d\xi_3  \! \mathrm{\int} \! \frac{d^4k}{(2\pi)^4} \frac{(-3!) i\eta_{\rho\sigma_1}}{(k^2 \!+\!\xi_1\!+\! \xi_2 \!+\!\xi_3 \!+\!i\epsilon)^4}   \nonumber \\
&& \times \frac{i\eta_{\rho_1\sigma}}{(k\!-\!p)^2\!+\!i\epsilon} \Big((g^{\rho\sigma} g^{{\mu_0}{\nu_0}})^{\mu\nu} (-k)_{\mu_0} (k-p)_{\nu_0}    \\
&&  - (g^{\rho{\mu_0}} g^{\sigma{\nu_0}} g^{\mu_1\nu_1})  p_\lambda  \{ (-k)_{\nu_1} (\Gamma_{{\nu_0}{\mu_1}{\mu_0}})^{\lambda\mu\nu} \nonumber \\
&& -(k-p)_{\nu_1} (\Gamma_{{\mu_0}{\mu_1}{\nu_0}})^{\lambda\mu\nu} + p_{\mu_1} (\Gamma_{{\nu_1}{\mu_0}{\nu_0}})^{\lambda\mu\nu}  \nonumber \\
&& - p_{\mu_0} (\Gamma_{{\mu_1}{\nu_0}{\nu_1}})^{\lambda\mu\nu} \} \Big)   \Big((g^{{\rho_1}{\sigma_1}} g^{{\alpha_0}{\beta_0}})^{\alpha\beta} (p-k)_{\alpha_0} k_{\beta_0}    \nonumber \\
&&  - (g^{{\rho_1}{\alpha_0}} g^{{\sigma_1}{\beta_0}} g^{\alpha_1\beta_1})  (-p)_{\lambda_1}  \{ (p-k)_{\beta_1} (\Gamma_{{\beta_0}{\alpha_1}{\alpha_0}})^{{\lambda_1}\alpha\beta} \nonumber \\
&& -k_{\beta_1} (\Gamma_{{\alpha_0}{\alpha_1}{\beta_0}})^{{\lambda_1}\alpha\beta} + (-p)_{\alpha_1} (\Gamma_{{\beta_1}{\alpha_0}{\beta_0}})^{{\lambda_1}\alpha\beta}  \nonumber \\
&& - (-p)_{\alpha_0} (\Gamma_{{\alpha_1}{\beta_0}{\beta_1}})^{{\lambda_1}\alpha\beta} \} \Big) \Big]_{\{\xi_1, \xi_2, \xi_3\} \to 0} \! + \!C_c^{\mu\nu\alpha\beta} \, .   \nonumber
\end{eqnarray}
After integral, one has
\begin{eqnarray}
\mathcal{T}_\mathrm{P}^{c} &=& (-1)(i \kappa)^{2}\frac{4i}{16\pi^2} \mathrm{\int}^1_{\!\!\! 0}  \!d x  (-3!)   \eta_{\rho\sigma_1} \eta_{\rho_1\sigma}  \\
&& \times \frac{1}{4!} \Big\{ \Big( W_g^{11} x(1-x) [x(1-x)p_{\alpha_0}^{} p_{\beta_0}^{} p_{\mu_0}^{} p_{\nu_0}^{} \nonumber \\
&& +\frac{p^4}{8} x(1\!-\!x) (\eta_{\beta_0 \mu_0}^{} \eta_{\alpha_0 \nu_0}^{}\!\!+\!\eta_{\alpha_0 \mu_0}^{} \eta_{\beta_0 \nu_0}^{}\!\!+\!\eta_{\alpha_0 \beta_0}^{} \eta_{\mu_0 \nu_0}^{})  \nonumber \\
&&+\frac{p^2}{2} ( x(1\!-\!x)(\eta_{\mu_0 \nu_0}^{} p_{\alpha_0}^{}p_{\beta_0}^{}\!\!+\!\eta_{\beta_0 \nu_0}^{} p_{\alpha_0}^{}p_{\mu_0}^{} \!\! +\!\eta_{\alpha_0 \mu_0}^{} p_{\nu_0}^{}p_{\beta_0}^{} \nonumber \\
&&+\eta_{\alpha_0 \beta_0}^{} p_{\mu_0}^{}p_{\nu_0}^{}) \!-\!x^2\eta_{\beta_0 \mu_0}^{} p_{\alpha_0}^{} p_{\nu_0}^{} \!\!-\!(1\!-\!x)^2 \eta_{\alpha_0 \nu_0}^{} p_{\beta_0}^{} p_{\mu_0}^{} )   ]  \nonumber \\
&& + W_g^{12} x (1\!-\!x) [((1\!-\!x)p_{\beta_1}^{} (\Gamma)^{\lambda_1} - p_y) p_{\lambda_1}^{} p_{\mu_0}^{}p_{\nu_0}^{} \nonumber \\
&& -\frac{p^2}{2} p_{\lambda_1}^{} (\eta_{\mu_0 \nu_0}^{} p_y + (\Gamma)^{\lambda_1} (x \eta_{\beta_1 \mu_0}^{} p_{\nu_0}^{} \nonumber \\
&& -(1-x)(\eta_{\mu_0 \nu_0}^{} p_{\beta_1}^{} + \eta_{\beta_1 \nu_0}^{} p_{\mu_0}^{})))]  \nonumber \\
&&+  W_g^{21} x (1\!-\!x) [(p_x-(1\!-\!x)p_{\nu_1}^{} (\Gamma)^{\lambda}) p_{\lambda}^{} p_{\beta_0}^{} p_{\alpha_0}^{} \nonumber \\
&& +\frac{p^2}{2} p_{\lambda}^{} (\eta_{\beta_0 \alpha_0}^{} p_x + (\Gamma)^{\lambda}(x \eta_{\nu_1 \beta_0}^{} p_{\alpha_0}^{} \nonumber \\
&& -(1-x)(\eta_{\beta_0 \alpha_0}^{} p_{\nu_1}^{} + \eta_{\nu_1 \alpha_0}^{} p_{\beta_0}^{})))] \nonumber \\
&&+ W_g^{22} p_{\lambda}^{} p_{\lambda_1}^{} (- p_x p_y + \frac{1-x}{2}(x p^2 \eta_{\beta_1 \nu_1}^{} (\Gamma)^{\lambda} (\Gamma)^{\lambda_1} \nonumber \\
&& + 2 p_x p_{\beta_1}^{} \! (\Gamma)^{\lambda_1} \!\!+\! 2 p_y p_{\nu_1}^{} \! (\Gamma)^{\lambda} \!\!-\! 2 (1\!-\!x)p_{\beta_1}^{} p_{\nu_1}^{} \! (\Gamma)^{\lambda} (\Gamma)^{\lambda_1} \! ) ) \! \Big)  \nonumber \\
&&\times \log\!\frac{1}{-p^2 x(1\!-\!x)} \Big \} \! + \! C_c^{\mu\nu\alpha\beta} \, ,   \nonumber
\end{eqnarray}
with
\begin{eqnarray}
W_g^{11} = (g^{\rho\sigma} g^{{\mu_0}{\nu_0}})^{\mu\nu} (g^{{\rho_1}{\sigma_1}} g^{{\alpha_0}{\beta_0}})^{\alpha\beta} \, , \nonumber
\end{eqnarray}
\begin{eqnarray}
W_g^{12} = - (g^{\rho\sigma} g^{{\mu_0}{\nu_0}})^{\mu\nu} (g^{{\rho_1}{\alpha_0}} g^{{\sigma_1}{\beta_0}} g^{\alpha_1\beta_1}) \, , \nonumber
\end{eqnarray}
\begin{eqnarray}
W_g^{21} = - (g^{\rho{\mu_0}} g^{\sigma{\nu_0}} g^{\mu_1\nu_1}) (g^{{\rho_1}{\sigma_1}} g^{{\alpha_0}{\beta_0}})^{\alpha\beta} \, , \nonumber
\end{eqnarray}
\begin{eqnarray}
W_g^{22} = (g^{\rho{\mu_0}} g^{\sigma{\nu_0}} g^{\mu_1\nu_1}) (g^{{\rho_1}{\alpha_0}} g^{{\sigma_1}{\beta_0}} g^{\alpha_1\beta_1})  \, , \nonumber
\end{eqnarray}
\begin{eqnarray}
(\Gamma)^{\lambda} = (\Gamma_{{\nu_0}{\mu_1}{\mu_0}})^{\lambda\mu\nu} + (\Gamma_{{\mu_0}{\mu_1}{\nu_0}})^{\lambda\mu\nu}  \, , \nonumber
\end{eqnarray}
\begin{eqnarray}
(\Gamma)^{\lambda_1} = (\Gamma_{{\beta_0}{\alpha_1}{\alpha_0}})^{{\lambda_1}\alpha\beta} + (\Gamma_{{\alpha_0}{\alpha_1}{\beta_0}})^{{\lambda_1}\alpha\beta}  \, , \nonumber
\end{eqnarray}
\begin{eqnarray}
p_x &=& p_{\nu_1} (\Gamma_{{\mu_0}{\mu_1}{\nu_0}})^{\lambda\mu\nu}  - p_{\mu_0} (\Gamma_{{\mu_1}{\nu_0}{\nu_1}})^{\lambda\mu\nu} \nonumber \\
&& + p_{\mu_1} (\Gamma_{{\nu_1}{\mu_0}{\nu_0}})^{\lambda\mu\nu}   \, , \nonumber
\end{eqnarray}
\begin{eqnarray}
p_y &=& p_{\beta_1} (\Gamma_{{\beta_0}{\alpha_1}{\alpha_0}})^{{\lambda_1}\alpha\beta} -p_{\alpha_1} (\Gamma_{{\beta_1}{\alpha_0}{\beta_0}})^{{\lambda_1}\alpha\beta}  \nonumber \\
&& +p_{\alpha_0} (\Gamma_{{\alpha_1}{\beta_0}{\beta_1}})^{{\lambda_1}\alpha\beta}   \, . \nonumber
\end{eqnarray}
After contraction, the result is
\begin{eqnarray}
\mathcal{T}_\mathrm{P}^{c} \!&=&\! (-1)(i \kappa)^{2}\frac{4i}{16\pi^2} \mathrm{\int}^1_{\!\!\! 0}  \!d x  (-\frac{1}{4})  \Big\{\frac{1}{4} [4 (4x^4 \!-\!8x^3 \!+\!2x^2    \\
&&+2x \!+\!1) p^{\mu} p^{\nu} p^{\alpha} p^{\beta} +p^2((8x^4 \!-\!16x^3 \!+\!4x^2 \!+\!4x\!-\!1) \nonumber  \\
&& \times (p^{\nu} p^{\beta} \eta^{\mu \alpha} \!+\!p^{\mu} p^{\beta} \eta^{\nu \alpha} +p^{\nu} p^{\alpha} \eta^{\mu \beta} \!+\!p^{\mu} p^{\alpha} \eta^{\nu \beta})  \nonumber \\
&&+2x(14x^3\!-\!24x^2\!+\!13x\!-\!4)p^{\mu} p^{\nu} \eta^{\alpha \beta} \!+\! 2 p^{\alpha} p^{\beta} \eta^{\mu \nu}  \nonumber \\
&& \times (14x^4\!-\!32x^3\!+\!25x^2\!-\!6x\!-\!1)) \!+\!p^4(2x(11x^3 \!-\! 22x^2  \nonumber  \\
&& +13x\!-\!2)(\eta^{\mu \alpha} \eta^{\nu \beta} \!+\! \eta^{\mu \beta} \eta^{\nu \alpha})\!+\!(12x^4 \!-\!24x^3\!+\!16x^2 \nonumber \\
&& -4x\!+\!1)\eta^{\mu \nu} \eta^{\alpha \beta} )] \log\!\frac{1}{-p^2 x(1\!-\!x)} \Big \}  + C_c^{\mu\nu\alpha\beta} \, .   \nonumber
\end{eqnarray}
The $\mu\nu \leftrightarrow \alpha\beta$ asymmetry involved at one-loop level in a particle propagation means that time reversal is not invariant in quantum gravity, i.e. an arrow of time at the microscopic level. The total one-loop physical amplitude $\mathcal{T}_\mathrm{P}^{1}$ is
\begin{eqnarray}
\mathcal{T}_\mathrm{P}^{1} &=& \mathcal{T}_\mathrm{P}^{a} + \mathcal{T}_\mathrm{P}^{b} +\mathcal{T}_\mathrm{P}^{c} \\
&=& (i \kappa)^{2}\frac{i}{16\pi^2} \mathrm{\int}^1_{\!\!\! 0}  \!d x  [\frac{1}{4} (11x^4 \!-\!22x^3 \!+\!3x^2 \!+\!8x \!+\!4) \nonumber  \\
&& \times p^{\mu} p^{\nu} p^{\alpha} p^{\beta} +\frac{p^2}{16}((22x^4 \!-\!44x^3 \!-\!x^2 \!+\!23x\!-\!6) \nonumber  \\
&& \times (p^{\nu} p^{\beta} \eta^{\mu \alpha} \!+\!p^{\mu} p^{\beta} \eta^{\nu \alpha} +p^{\nu} p^{\alpha} \eta^{\mu \beta} \!+\!p^{\mu} p^{\alpha} \eta^{\nu \beta})  \nonumber \\
&&+(52x^4-72x^3+37x^2-25x+2)p^{\mu} p^{\nu} \eta^{\alpha \beta}    \nonumber \\
&&+ (52x^4\!-\!136x^3\!+\!133x^2\!-\!41x\!-\!6)p^{\alpha} p^{\beta} \eta^{\mu \nu})  \!+\! \frac{p^4}{32} \nonumber  \\
&&\times ((91x^4\!\!-\!182x^3\!\!+\!69x^2\!\!+\!22x\!-\!3)(\eta^{\mu \alpha} \eta^{\nu \beta} \!\!+\! \eta^{\mu \beta} \eta^{\nu \alpha}) \nonumber \\
&& +(-19x^4\!\!+\!38x^3\!\!+\!25x^2\!\!-\!44x\!+\!7)\eta^{\mu \nu} \eta^{\alpha \beta} )] \log\!\frac{\mu^2}{-p^2}   \, ,   \nonumber
\end{eqnarray}
with an energy scale $-p^2 = \mu^2$ adopted.

\subsection{$n-$loop with overlapping divergences}

Here we give a brief discussion about $n-$loop graviton with overlapping/nested divergences. To a closed graviton loop, the superficial degree of divergence by power counting is 4. Generally, for $n-$nested loop of graviton, the superficial degree of divergence is up to a power of $2n+2$. To a term of $2n$ power divergence, the corresponding physical transition amplitude $\mathcal{T}_\mathrm{P}^{t2n}$ can be written as (see the Appendix B)
\begin{eqnarray}
\mathcal{T}_\mathrm{P}^{t2n} = A \frac{\Delta^n}{n!} \log\! \mid\! \Delta\!\mid  + C \, .
\end{eqnarray}
For $n-$nested loop of graviton with divergences up to $2(n\!+\!1)$-th power, the physical transition amplitude $\mathcal{T}_\mathrm{P}^{\mathrm{total}}$ can be written as
\begin{eqnarray}
\mathcal{T}_\mathrm{P}^{\mathrm{total}} &=& \mathcal{T}_\mathrm{P}^{t2(n\!+\!1)} + \mathcal{T}_\mathrm{P}^{t2n} + \cdots + \mathcal{T}_\mathrm{P}^{t2}\\
 &&+\mathcal{T}_\mathrm{P} (\log) + \mathcal{T}_\mathrm{P} (\mathrm{finite})\, , \nonumber
\end{eqnarray}
with $\mathcal{T}_\mathrm{P} (\log)$ being $\log$-divergence contributions, and $\mathcal{T}_\mathrm{P} (\mathrm{finite})$ being originally finite terms. The $n-$nested graviton loop can be described in UV-free scheme.

\begin{figure}[htbp!]
\includegraphics[width=0.14\textwidth]{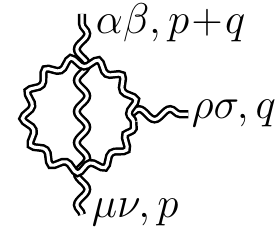} \vspace*{-1ex}
\caption{A two-loop diagram of three-graviton vertex.}
\label{two-loop}
\end{figure}

Let's look at a specific two-loop correction of three-graviton vertex, as shown in Fig. \ref{two-loop}. The sextic divergence (to the power of six) is involved in this two-loop process, and there are 108 vertex-product terms. The physical transition amplitude $\mathcal{T}_\mathrm{P}^{V}$ of this process is
\begin{eqnarray}
\mathcal{T}_\mathrm{P}^{V}\!\!\! & = & \!\! \bigg[\! \int \! (d\xi)^4  \frac{\partial^4\mathcal{T}_\mathrm{F}^V (\xi)}{\partial\xi^4}   \!\bigg]_{\xi \to 0} \! + \! C^{\mu\nu\alpha\beta\rho\sigma}       \\
\!\!\!&=&\!\!\bigg[ (2 i)^3 \kappa^5 \!\! \int \! (d\xi)^4 \! \mathrm{\int} \!\! \frac{d^4k_A}{(2\pi)^4} \! \frac{d^4k_B}{(2\pi)^4} \frac{4! i\Pi_{\alpha_2\beta_2\mu_4\nu_4}}{2^4(k_A^2 \!-\!\xi \!+\! i\epsilon)^5} \nonumber \\
&&\!\! \times  \frac{ i\Pi_{\mu_3\nu_3\alpha_3\beta_3} }{k_N^2 \!+\! i\epsilon }  \frac{ i\Pi_{\rho_3\sigma_3\mu_2\nu_2} }{k_B^2 \!+\! i\epsilon } \frac{ i\Pi_{\alpha_4\beta_4\rho_2\sigma_2}}{(k_B\!-\!q)^2 \!+\! i\epsilon } \nonumber \\
&&\!\! \times \Big(\!V^{\mu_3\nu_3\mu_4\nu_4|\lambda_1\mu\nu \lambda_2\mu_2\nu_2}  p_{\lambda_1} \!(k_B\!)_{\!\lambda_2} \!\!+\!V^{\mu_2\nu_2\mu_4\nu_4|\lambda_1\mu\nu \lambda_3\mu_3\nu_3} \nonumber\\
&&\!\! \times p_{\lambda_1} \!(-k_N\!)_{\!\lambda_3} \!\!+\!V^{\mu_2\nu_2\mu_3\nu_3|\lambda_1\mu\nu \lambda_4\mu_4\nu_4}  p_{\lambda_1} \!(k_A)_{\!\lambda_4} \nonumber\\
&&\!\!+V^{\mu\nu \mu_4\nu_4|\lambda_2\mu_2\nu_2 \lambda_3\mu_3\nu_3}  (k_B\!)_{\!\lambda_2} \!(\!-k_N\!)_{\!\lambda_3} \nonumber \\
&&\!\!+\!V^{\!\mu\nu\mu_3\nu_3|\lambda_2\mu_2\nu_2 \lambda_4\mu_4\nu_4} (k_B\!)_{\!\lambda_2} \!(k_A\!)_{\!\lambda_4} \!\!+\!\!V^{\!\mu\nu\mu_2\nu_2|\lambda_3\mu_3\nu_3 \lambda_4\mu_4\nu_4} \nonumber \\
&&\!\! \times \!(\!-k_N\!)_{\!\lambda_3} \!(k_A)_{\!\lambda_4} \!\Big) \Big(\!V^{\alpha_3\beta_3\alpha_4\beta_4|\theta_1\alpha\beta \theta_2\alpha_2\beta_2} (-l)_{\!\theta_1} \!(-k_A)_{\!\theta_2}  \nonumber \\
&&\!\! + V^{\alpha_2\beta_2\alpha_4\beta_4|\theta_1\alpha\beta \theta_3\alpha_3\beta_3} (-l)_{\!\theta_1} (k_N)_{\!\theta_3} \nonumber\\
&&\!\!+V^{\alpha_2\beta_2\alpha_3\beta_3|\theta_1\mu\nu \theta_4\alpha_4\beta_4} (-l)_{\!\theta_1} (q\!-\!k_B)_{\!\theta_4} \nonumber\\
&&\!\!+V^{\alpha\beta \alpha_4\beta_4|\theta_2\alpha_2\beta_2 \theta_3\alpha_3\beta_3} (-k_A)_{\!\theta_2} (k_N)_{\!\theta_3} \nonumber \\
&&\!\!+V^{\alpha\beta\mu_3\nu_3|\theta_2\alpha_2\beta_2 \theta_4\alpha_4\beta_4} (-k_A)_{\!\theta_2} (q\!-\!k_B)_{\!\theta_4} \nonumber \\
&&\!\!+\!V^{\alpha\beta\alpha_2\beta_2|\theta_3\alpha_3\beta_3 \theta_4\alpha_4\beta_4} \!(k_N)_{\!\theta_3} \!(q\!-\!k_B)_{\!\theta_4} \!\!\Big) \! \Big(\!V^{\rho_3\sigma_3|\delta_1\rho\sigma \delta_2\rho_2\sigma_2}\nonumber \\
&&\!\! \times q_{\delta_1}^{} \!(k_B\!-\!q)_{\!\delta_2}^{} \!\!+\!V^{\rho_2\sigma_2|\delta_1\rho\sigma \delta_3\rho_3\sigma_3} \!q_{\delta_1}^{} \!(\!-k_B\!)_{\!\delta_3}^{}   \nonumber \\
&&\!\!+V^{\rho\sigma|\delta_2\rho_2\sigma_2 \delta_3\rho_3\sigma_3} (k_B\!-\!q)_{\delta_2}^{} (-k_B)_{\delta_3}^{} \! \Big)  \!\bigg]_{\xi \to 0} \!+\! C^{\mu\nu\alpha\beta\rho\sigma} \nonumber \, ,
\end{eqnarray}
with $k_N \!=\! k_A\!+\!k_B\!+\!p$, $l \!=\! p\!+\!q$. After the integral and contraction, the result will be obtained, which can be written as
\begin{eqnarray}
\mathcal{T}_\mathrm{P}^{V} \!\!&=&\! (2 i)^3 \kappa^5 (\frac{i}{16 \pi^2})^2 \!\mathrm{\int}_{\!\!0}^1 \!dx\! \!\mathrm{\int}_{\!\!0}^1 \!dy\! \!\mathrm{\int}_{\!\!0}^1 \!dz  \frac{z}{2^4(1-z)^4}   \\
&&\!\! \times \!\bigg\{ A_3 \frac{\Delta^3_0}{3!} + A_2 \frac{\Delta^2_0}{2!} + A_1 \Delta^{}_0 + A_0 \!\bigg\} \log\!\frac{1}{\Delta^{}_0} \!+\! C^{\mu\nu\alpha\beta\rho\sigma} \nonumber \, .
\end{eqnarray}
Here $\Delta^{}_0$ is $\Delta^{}_0 = b^2-ac$, with $a= z +(1-z)x(x-1)$, $b= y z q +(1-z)x(x-1)p$, $c= y z q^2 +(1-z)x(x-1)p^2$. $A_3$, $A_2$, $A_1$, $A_0$ are coefficients related to sextic, quartic, quadratic, logarithmic divergence inputs respectively. In this paper, we focus on the feasibility of the method, and here the high-power coefficient $A_3$ is evaluated as an illustration (the parameters $A_0$, $A_1$, and $A_2$ have large numbers of terms, and the result of $A_0$ is listed in the Supplement as an expample), with
\begin{eqnarray}
A_3 &=& \frac{z-1}{64 a^8} \big( [440 a^2+a (1564 x^2+1300 x+23) (z-1) \\
&& +4 (281 x^4-562 x^3+683 x^2-402 x+273) (z-1)^2] \nonumber \\
&& \times \eta ^{\mu \nu } (\eta ^{\alpha \rho } \eta ^{\beta \sigma }+\eta ^{\alpha \sigma } \eta ^{\beta \rho }) + [744 a^2+a (1932 x^2+44 x \nonumber \\
&& +1203) (z-1)+4 (297 x^4\!-594 x^3\!+1563 x^2\!-1266 x \nonumber \\
&& +673) (z-1)^2] \eta ^{\rho \sigma } (\eta ^{\alpha \nu } \eta ^{\beta \mu }+\eta ^{\alpha \mu } \eta ^{\beta \nu }) + [440 a^2 \nonumber \\
&& +a (1564 x^2-1100 x+2423) (z-1)+4 (281 x^4 \nonumber \\
&& -562 x^3+683 x^2-402 x+273) (z-1)^2] \eta ^{\alpha \beta } (\eta ^{\mu \rho } \eta ^{\nu \sigma }   \nonumber \\
&&  +\eta ^{\mu \sigma } \eta ^{\nu \rho }) + [1032 a^2+a (3396 x^2-3020 x+801) \nonumber \\
&& \times (z-1)+4 (591 x^4\!-\!1182 x^3\!+\!1101 x^2\!-\!510 x\!+\!215) \nonumber \\
&& \times (z-1)^2] (\eta ^{\alpha \rho } \eta ^{\beta \nu } \eta ^{\mu \sigma }+\eta ^{\alpha \nu } \eta ^{\beta \rho } \eta ^{\mu \sigma }+\eta ^{\alpha \nu } \eta ^{\beta \sigma } \eta ^{\mu \rho } \nonumber \\
&& +\eta ^{\alpha \sigma } \eta ^{\beta \nu } \eta ^{\mu \rho } \!+\!\eta ^{\alpha \rho } \eta ^{\beta \mu } \eta ^{\nu \sigma }\!+\!\eta ^{\alpha \mu } \eta ^{\beta \rho } \eta ^{\nu \sigma }+\eta ^{\alpha \mu } \eta ^{\beta \sigma } \eta ^{\nu \rho } \nonumber \\
&& +\eta ^{\alpha \sigma } \eta ^{\beta \mu } \eta ^{\nu \rho }) \!+\! [1696 a^2+a (4844 x^2+848 x+4147) \nonumber \\
&& \times (z-1)+4 (787 x^4-1574 x^3+2521 x^2-1734 x \nonumber \\
&& +795) (z-1)^2] \eta ^{\alpha \beta } \eta ^{\mu \nu } \eta ^{\rho \sigma }  \big) \nonumber \, .
\end{eqnarray}

\section{Conclusion and Discussion}

In this paper, we have given a try to describe possible quantum behavior of Einstein gravity, and focused on graviton loops in the UV-free scheme. To the problem of the non-renormalizability of Einstein gravity, it seems to be an issue of how to properly describe UV regions of loops. Rather than an infinite number of unfixed counterterms (higher and higher power of the Riemann curvature tensor) required to cure UV divergences of graviton loops in the procedure of renormalization, there is no requirement for high power of the Riemann tensor (only a basic term of the curvature scalar $R$ is involved) for graviton loops in the UV-free scheme. The one-loop result of graviton propagation indicates that time reversal is not invariant in quantum gravity. The UV-free scheme seems to be effective for graviton loops, i.e. it is possible to incorporate gravitation into the framework of quantum field theory. This indicates that both loops of the renormalizable field and the non-renormalizable field (SM+Gravity) can be described in a unified way of UV-free scheme, i.e. an alternative method of loops for four fundamental interactions. Moreover, it is an effective perturbative description of quantum gravity below the Planck scale, with possible quantum gravity at the Planck scale unclear at present. We look forward to more explorations in the future.

\acknowledgments

This work was partly supported by the open project of the theoretical physics academic exchange platform of Chongqing University.

\appendix

\section{Feynman rules for gravitons}  \label{add-ex}

Here the Feynman rules for gravitons in a weak field expansion around Minkowski spacetime with the coordinate invariant volume transformation are listed below. The propagator of graviton is set by the quadratic term of $ h_{\mu \nu}$ in the expansion. The results ($\zeta = 2$) of the propagators of graviton and ghost, the vertexes of multi-graviton and graviton-ghost are
\begin{figure}[h]
 \centering
  \begin{tikzpicture}
    \node
    (g-p)
        {
          \includegraphics[width=0.17\textwidth]{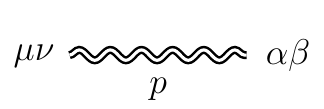}
        };
        \node
            [right=-0.0cm of g-p]
            {
              $= \frac{i\Pi_{\mu\nu\alpha\beta}/2}{p^2 +i\epsilon}$ .
            };
  \end{tikzpicture}
\end{figure}
\begin{figure}[h]
 \centering
  \begin{tikzpicture}
    \node
    (g-p)
        {
          \includegraphics[width=0.165\textwidth]{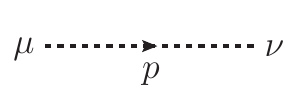}
        };
        \node
            [right=-0.0cm of g-p]
            {
              =  \large $\frac{i \eta_{\mu\nu}}{p^2 + i \epsilon}$ .
            };
  \end{tikzpicture}
\end{figure}\\
\quad \\
\begin{figure}[h!]
 \includegraphics[width=0.2\textwidth]{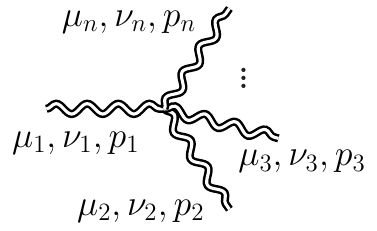}\\
  $= 2 i \kappa^{n-2} \big[ V^{\mu_3\nu_3 \cdots \mu_n\nu_n|\lambda_1\mu_1\nu_1 \lambda_2\mu_2\nu_2} (p_1)_{\lambda_1} (p_2)_{\lambda_2}$ \\ + permutations$\big]$ , \\
 \quad \\
 \includegraphics[width=0.18\textwidth]{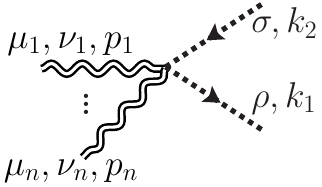} \\
       = $ i \kappa^{n} V_g^{\mu_1\nu_1 \cdots \mu_n\nu_n|\rho\sigma} $ .
\end{figure} \\
\quad \\
\quad \\
The multi-graviton parameter is $V^{\mu_3\nu_3 \cdots \mu_n\nu_n|\lambda_1\mu_1\nu_1 \lambda_2\mu_2\nu_2}$ $ = \big(g^{\mu\nu} g^{\alpha\beta} g^{\rho\sigma}\big)^{\mu_3\nu_3 \cdots \mu_n\nu_n} [ (\Gamma_{\alpha\mu\rho})^{\lambda_1\mu_1\nu_1} (\Gamma_{\sigma\nu\beta})^{\lambda_2\mu_2\nu_2} $ \\
  $ - \frac{1}{2} (\Gamma_{\alpha\mu\nu})^{\lambda_1\mu_1\nu_1} (\Gamma_{\rho\beta\sigma})^{\lambda_2\mu_2\nu_2} -\frac{1}{2}(\Gamma_{\rho\beta\sigma}\!)^{\lambda_1\mu_1\nu_1} (\Gamma_{\alpha\mu\nu}\!)^{\lambda_2\mu_2\nu_2} $ \\
  $ -\frac{1}{2} (\Gamma_{\mu\alpha\beta}\!)^{\lambda_1\mu_1\nu_1} (\Gamma_{\nu\rho\sigma}\!)^{\lambda_2\mu_2\nu_2} ]$.
Here some terms are $(g^{\mu\nu})$ \\
$=\! \eta^{\mu\nu}$,  $(g^{\mu\nu})^{\mu_i\nu_i}$ $= - \frac{1}{2} (\eta^{\mu\mu_i}\eta^{\nu\nu_i} + \eta^{\mu\nu_i}\eta^{\nu\mu_i})$, $(\Gamma_{\mu\alpha\beta})^{\lambda_i\mu_i\nu_i}$ \\ $=\!\frac{1}{2}[\delta^{\lambda_i}_\alpha \!I_{\beta\mu}^{\mu_i\nu_i} \!+\! \delta^{\lambda_i}_\beta \!I_{\alpha\mu}^{\mu_i\nu_i} \!-\! \delta^{\lambda_i}_\mu \!I_{\alpha\beta}^{\mu_i\nu_i}]$ with $I_{\alpha\beta}^{\mu\nu} \!=\!\frac{1}{2} (\delta^{\mu}_\alpha \delta^{\nu}_\beta \!+\! \delta^{\mu}_\beta \delta^{\nu}_\alpha)$. All momenta are considered to be inwards. Permutations of the graviton field terms (in the form of parameters \{$\mu_i, \nu_i, p_i$\}) are proceeded, which are symmetric in the graviton fields. The graviton-ghost parameter is $V_g^{\mu_1\nu_1 \cdots \mu_n\nu_n|\rho\sigma}$  $= (g^{\rho\sigma} g^{\alpha\beta})^{\mu_1\nu_1 \cdots \mu_n\nu_n} (k_1)_\alpha (k_2)_\beta $
  $- \{(g^{\rho\alpha} g^{\sigma\beta} g^{\mu\nu})^{\mu_2\nu_2 \cdots \mu_n\nu_n}  (p_1)_\lambda \big( (k_1)_\nu (\Gamma_{\beta\mu\alpha})^{\lambda\mu_1\nu_1}$
  $-(k_2)_\nu (\Gamma_{\alpha\mu\beta})^{\lambda\mu_1\nu_1} \!+\! (p_1)_\mu (\Gamma_{\nu\alpha\beta})^{\lambda\mu_1\nu_1} \!-\! (p_1)_\alpha (\Gamma_{\mu\beta\nu})^{\lambda\mu_1\nu_1}  \big)$ \\
  + permutations $\}$ $- \{ (g^{\rho\alpha} g^{\sigma\beta} g^{\mu\nu} g^{\lambda\tau})^{\mu_3\nu_3 \cdots \mu_n\nu_n}$ \\
  $\times \frac{1}{2}[ (\Gamma_{\mu\alpha\lambda})^{\lambda_1\mu_1\nu_1} (\Gamma_{\nu\beta\tau})^{\lambda_2\mu_2\nu_2}  - (\Gamma_{\mu\alpha\beta})^{\lambda_1\mu_1\nu_1} (\Gamma_{\nu\lambda\tau})^{\lambda_2\mu_2\nu_2}$ $+(\Gamma_{\alpha\mu\lambda})^{\lambda_1\mu_1\nu_1} (\Gamma_{\beta\nu\tau})^{\lambda_2\mu_2\nu_2} $
  $+ (\{\!\lambda_1,\mu_1,\nu_1\!\} \!\leftrightarrow\! \{\!\lambda_2,\mu_2,\nu_2 \!\})]$  $\times (p_1)_{\lambda_1} (p_2)_{\lambda_2}$ + permutations$\} $.

\section{$\xi-$dependent choice}  \label{xi-d-choice}

The UV-free scheme is a way to describe the physical transition amplitude $\mathcal{T}_\mathrm{P}$ of both tree-level and loop-level processes (the tree-level application is a trivial case, see Ref. \cite{Jia:2023dub}). For loops with power-law divergences, $\xi-$independent terms appear in the primary antiderivative before taking the limit $\xi \to 0$. The $\xi-$dependent choice is that the primary antiderivative consists of $\xi-$dependent terms, with $\xi-$independent terms absorbed into the boundary constant $C$. For a quadratic divergence with $\xi-$dependent choice, the corresponding physical transition amplitude $\mathcal{T}_\mathrm{P}^{t2}$ can be written as
\begin{eqnarray}
\mathcal{T}_\mathrm{P}^{t2} &=& A \bigg[ (\xi + \Delta)(\log\! \mid\!\xi \!+\! \Delta\!\mid -1) \bigg]_{\xi \to 0} + C_1  \\
&=& A \bigg[ (\xi + \Delta)\log\! \mid\!\xi \!+\! \Delta\!\mid - \xi \bigg]_{\xi \to 0} + C \, , \nonumber
\end{eqnarray}
with $A$ (a coefficient) and $\Delta$ being $\xi-$independent. For a quartic divergence with this choice, the corresponding physical transition amplitude $\mathcal{T}_\mathrm{P}^{t4}$ can be written as
\begin{eqnarray}
\mathcal{T}_\mathrm{P}^{t4} &=& A \bigg[ \frac{(\xi + \Delta)^2}{2}(\log\! \mid\!\xi \!+\! \Delta\!\mid -\frac{3}{2}) \bigg]_{\xi \to 0} + C_1  \\
&=& A \bigg[ \frac{(\xi + \Delta)^2}{2} \log\! \mid\!\xi \!+\! \Delta\!\mid - \frac{3}{4}(\xi^2 + 2 \xi \Delta) \bigg]_{\xi \to 0} + C \, . \nonumber
\end{eqnarray}
Taking the limit $\xi \to 0$, the relic $\log$ term becomes the final primary antiderivative of a power-law divergence. For loops with high power divergences (e.g. graviton loops with overlapping/nested divergences), i.e. to a power of $2n$ ($n \geq 1$), the corresponding physical transition amplitude $\mathcal{T}_\mathrm{P}^{t2n}$ with $\xi-$dependent choice can be written as
\begin{eqnarray}
\mathcal{T}_\mathrm{P}^{t2n} &=& A \bigg[ \frac{(\xi + \Delta)^n}{n!}(\log\! \mid\!\xi \!+\! \Delta\!\mid -(\sum_{l=1}^n \frac{1}{l})) \bigg]_{\xi \to 0} + C_1   \nonumber \\
&=& A \frac{\Delta^n}{n!} \log\! \mid\! \Delta\!\mid  + C \, .
\end{eqnarray}
The logarithmic expression describes the local relative evolution with renormalization conditions (or physical normalization conditions) adopted, i.e. $C$ given by typical reference energy points. In $\xi-$dependent choice, the primary antiderivative is well-defined for both tree-level and loop-level (include the case of loop finite, loop log and power-law divergences) processes. For instance, the physical transition amplitude of QED vacuum polarization (fermion loop) with this choice is
\begin{eqnarray}
\mathcal{T}_\mathrm{P}^{\mu \nu} \! &=& \!    -\frac{i e^2}{2 \pi^2}  \! \mathrm{\int}_{\! \! \! 0}^1 d x  (p^\mu p^\nu -g^{\mu \nu} p^2)x(1-x) \\
&& \times  \log(m^2-p^2x(1-x))  +   C^{\mu \nu} \, ,              \nonumber
\end{eqnarray}
with the Ward identity automatically preserved by the primary antiderivative.


\end{document}